\documentclass
  {article}
\usepackage{graphicx}

\title{Remarkable long-range-systematic in the binding energies of
$\alpha$-nuclei II}
\author{D.H.E. Gross\\
Hahn-Meitner Institute and Freie Universit{\"a}t Berlin,\\
Fachbereich Physik.\\ Glienickerstr. 100\\ 14109 Berlin, Germany}
\begin{document}
\maketitle
\begin{abstract}
In this letter I present further data that show the remarkable evidence for
the existence of an $\alpha$-cluster stucture in the ground states of
even-even $N=Z$ nuclei. Such a remarkable systematic was observed 20 years
ago in \cite{gross62} for these nuclei at $A\le 72$ and is extended here up
to $A=100$.
\end{abstract}

\section{Introduction}
Twenty years ago we published a remarkable systematic of nuclear binding
energies \cite{gross62}. The two-nucleon separation energies of
$\alpha$-nuclei where found to be approximately constant and equal to the
2-proton, 2-neutron and PN-separation energies of the $^4He$-nucleus. At
these days this systematics could only be followed up to $^{72}Kr$. In the
past 20 years the table of nuclear binding energies was considerably
extended\footnote{Nuclear Structure and Decay Data, National Nuclear Data
Center, Brookhaven National Laboratory, Upton, NY 11973-5000;
http://www.nndc.bnl.gov/nndc/nndcnsdd.html.}: It is now interesting how far
this systematic can be followed also in these new data. This is indeed well
possible as will be shown in this publication.

\begin{figure}
\includegraphics*[bb =12 2 456 301, angle=0, width=12cm,
clip=true]{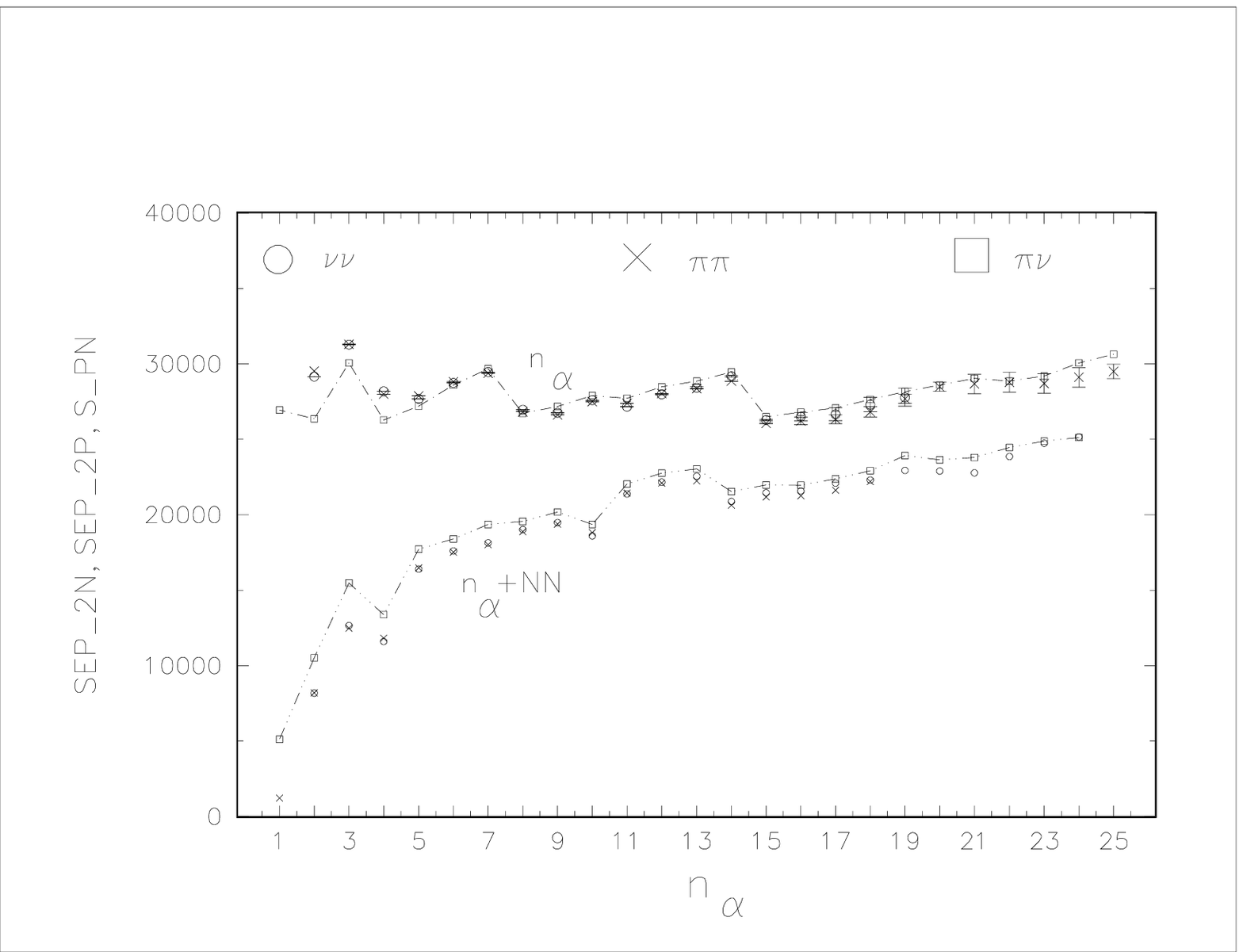}
\caption{Nuclear two-nucleon separation energies $S^{nuclear}_{2N}$,
corrected for Coulomb energies, in keV, following eq.\ref{coul}, with
$2N=\pi,\nu$ (neutron-proton pair), $2N=\nu,\nu$ (neutron-neutron pair),
and $2N=\pi,\pi$ (proton-proton pair) as function of the original number
$n_\alpha$ of $\alpha$'s. The upper curve is for  an
$n_\alpha$-mother-nucleus, the lower one is for a mother-nucleus with
$n_\alpha$ $\alpha$-particles plus one additional nucleon pair which is
then removed.
  \label{2nucleon}}
\end{figure}

The nuclear separation energies are defined by the difference of the
binding energies of initial nucleus minus the sum of that of the final
nuclei. I correct it in the most simple way for the different Coulomb
energies according to the trivial formula \ref{coul} \cite{gross62}:
\begin{eqnarray}
S^{nuclear}_{2N}&=&(B_i+U^C_i)-(B_f+U^C_f)-U^C_{2N}\\
U^C&=&\frac{3}{5}(e^2/r_0)Z(Z-1)/A^{1/3}, ~~~~r_0=1.25\mbox{  fm},
\label{coul}\end{eqnarray} where $B_i$ and $B_f$ are the experimental
binding energies of the initial and final nucleus respectively, $2N$
corresponds to any pair of nucleons [proton-proton ($\pi,\pi$),
neutron-proton ($\nu,\pi$), and neutron-neutron ($\nu,\nu$) pair]. As can
be clearly seen in figure \ref{2nucleon}, the removal of a 2N-pair costs
about {\em the same energy as for a bare $\alpha$-particle} $\approx
30$MeV. In contrast, the removal of an additional $\pi\pi$, $\nu\nu$ or
$\pi\nu$ pair costs much less energy. Moreover, this strongly depends on
the size the mother nucleus, resp. the number of $\alpha$-particles
surrounding this pair.

\begin{figure}
\includegraphics*[bb =18 10 451 293, angle=0, width=12cm,
clip=true]{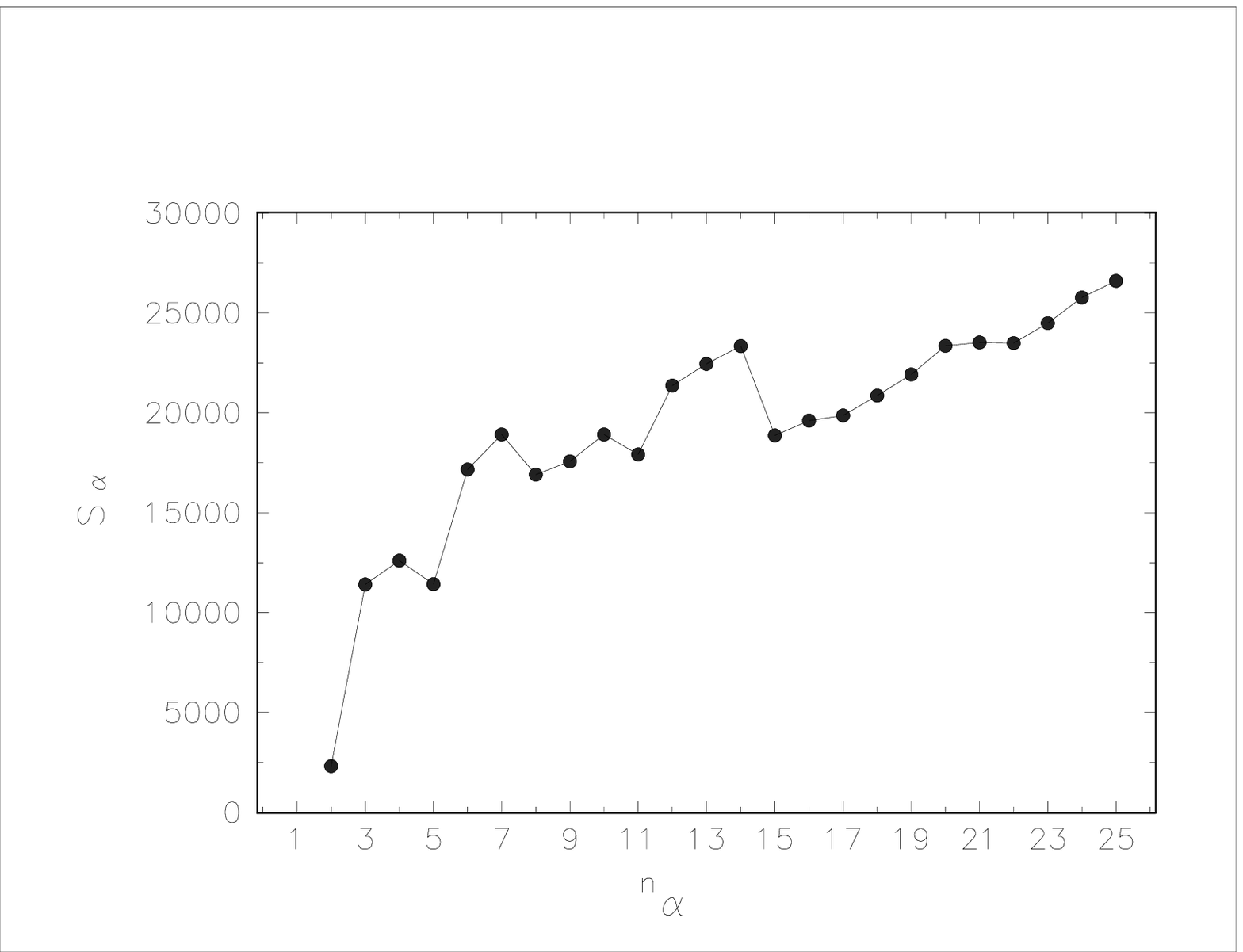}
\caption{The nuclear $\alpha$-separation energies $ S^{nuclear}_\alpha $
for $n_\alpha$-nuclei, corrected for Coulomb energies following
eq.\ref{coul}, in keV. There are peaks at $n_\alpha=4,7,14$ which are
followed by breaks. \label{ALPHA}}
\end{figure}

Figure \ref{ALPHA} show the separation energies of an $\alpha$-particle out
of a $n_\alpha$-nucleus. The peaks at $n_\alpha=4,7,14$, and the
disappearance of peaks above $n_\alpha=14$, remind of the special stable
$3$-dim clusters of closed packed spheres at $n=4,7,13,55$. The latter ones
with icosaeder structure are observed also for  atomic clusters of the
heavier noble gases like Xe$^n$. For the lighter ones like Ar clusters the
shift from $13$ to $14$ is well known.
\section{Acknowledgement}
I am grateful to W.v.Oertzen for pointing at the new and far more extended
table of nuclear binding energies.

\begin{thebibliography}{1}

\bibitem{gross62}
D.H.E. Gross and M.C. Nemes.
\newblock Remarkable long- range systematics in the binding energies of
  $\alpha$ nuclei.
\newblock {\em Phys. Lett}, B~130:131, 1983.

\end{thebibliography}

\end{document}